\documentclass{article}
\usepackage{spconf,amsmath,graphicx,hyperref}
\usepackage{amssymb,amsfonts}
\usepackage{amsthm}
\usepackage{algorithm}
\usepackage{algpseudocode}
\usepackage[table]{xcolor}
\usepackage{xcolor}
\usepackage{comment}
\usepackage{rotating}
\usepackage{booktabs}
\usepackage{textcomp} 
\usepackage{breqn}
\usepackage{siunitx}
\usepackage{musicography}

\newcommand{\stddev}[1]{{\color{gray}\footnotesize\,#1}}

\title{Single-Step Controllable \\ Music Bandwidth extension with Flow Matching
\thanks{%
\protect\scriptsize\normalfont
Copyright 2026 IEEE. Published in ICASSP 2026 - 2026 IEEE International Conference on Acoustics, Speech and Signal Processing (ICASSP), scheduled for 3-8 May 2026 in Barcelona, Spain. Personal use of this material is permitted. However, permission to reprint/republish this material for advertising or promotional purposes or for creating new collective works for resale or redistribution to servers or lists, or to reuse any copyrighted component of this work in other works, must be obtained from the IEEE. Contact: Manager, Copyrights and Permissions / IEEE Service Center / 445 Hoes Lane / P.O. Box 1331 / Piscataway, NJ 08855-1331, USA. Telephone: + Intl. 908-562-3966.}
}

\name{Carlos Hernandez-Olivan\sthanks{Work was conducted during an internship at Universal Music Group.} \qquad Hendrik Vincent Koops \qquad Hao Hao Tan \qquad Elio Quinton}

\address{Music and Audio Machine Learning Lab, Universal Music Group}

\begin{document}
%
\maketitle
\begin{abstract}
Audio restoration consists in inverting degradations of a digital audio signal to recover what would have been the pristine quality signal before the degradation occurred.
This is valuable in contexts such as archives of music recordings, particularly those of precious historical value, for which a clean version may have been lost or simply does not exist.
Recent work applied generative models to audio restoration, showing promising improvement over previous methods, and opening the door to the ability to perform restoration operations that were not possible before.
However, making these models finely controllable remains a challenge. In this paper, we propose an extension of \textsc{FLowHigh} and introduce the Dynamic Spectral Contour (\textsc{dsc}) as a control signal for bandwidth extension via classifier-free guidance.
Our experiments show competitive model performance, and indicate that \textsc{dsc} is a promising feature to support fine-grained conditioning.
\end{abstract}

\begin{figure*}[t]
    \centering
    \includegraphics[width=0.8\linewidth,trim={2.7cm 13cm 7.8cm 4.9cm},clip]{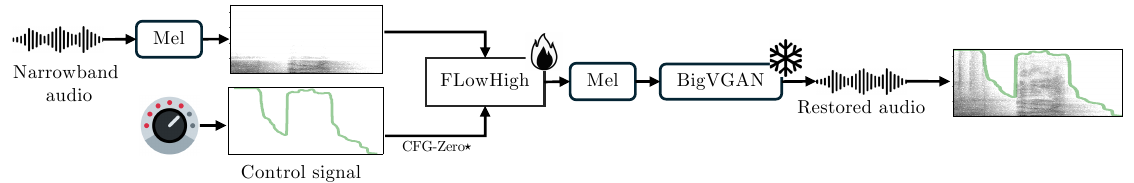}
    \caption{Proposed Flow Matching method with audio controls for music restoration.}
    \label{fig:general}
\end{figure*}

\section{Introduction}
\label{sec:intro}

Degradations in music recording may be introduced in countless ways such as poor recording conditions, legacy recording devices, lossy data compression, or degradation of the storage media like vinyl or magnetic tape. The presence of such degradations can make music recordings unusable, and can be an obstacle to the preservation of assets of particular historical value. 
Music restoration systems aim to improve audio quality of a degraded or low quality recording by reconstructing the missing audio information and/or removing artifacts. They hold the promise of allowing the salvage of recordings that otherwise could not be. For example, Moliner et al. use generative adversarial networks (\textsc{gan}s) for historical music bandwidth extension \cite{moliner2022behm}.

Recently, generative models have shown promising results in the broad context of audio restoration. For example, diffusion models trained with clean data were used for piano and vocal restoration \cite{cqtdiff,vrdmg}. 
However, diffusion models require many sampling steps to produce high-quality samples, impeding efficient and fast inference~\cite{10819705}. Flow-matching\cite{Lipman2023FlowMatching,Chen2018NeuralODE,Liu2023FlowStraightFast} is poised to allow single-step generation, which may yield faster generation.
\textsc{FLowHigh} is one such method proposed for speech super-resolution \cite{flowhigh}. 

As with most generative tasks, controllability of audio restoration models is an important property for their practical use. 
Although most audio restoration methods cited above do not allow for user control, recent work proposed using text prompts to guide the reconstruction \cite{sonicmaster}. 
To this day, the range of control signals explored in the audio and music generation space is comparatively larger. 
Recent work in music or foley sound generation introduces time-varying attributes to change audio features such as loudness, pitch and spectral centroid \cite{videocontrol,WuDWB24,sketch2sound}, or leverages multimodal controls \cite{foley}.
 
In this work we focus our attention on the task of generative bandwidth extension, which aims at reconstructing high frequency content that is missing from the original recording. 
With the view of designing tools for advanced and professional users, our objective is to investigate fine-grained controllability. Our hope is that such tool would empower producers and engineers alike to make the minute adjustments required for each recording, and perhaps even use the restoration creatively. 
Language, and by proxy text prompts, has proven to be a valuable control signal for generative models, but it is limited by what is possible with a language description. As an alternative that is more directly connected to the task of bandwidth extension and significantly less computationally intensive, we investigate using audio features as a fine-grained control signal.

\textbf{Contributions.}
Our contributions are the following:
(1) We transpose the \textsc{FLowHigh} framework to the music domain, to allow single-step generative restoration. (2) We extend \textsc{FLowHigh} with audio feature conditioning using \textsc{cfg-zero$\star$}. (3) We introduce a novel audio feature, Dynamic Spectral Contour (\textsc{dsc}), specifically designed to guide bandwidth extension. 

\section{Method}
\label{sec:method}

\subsection{Model Setup}

Our approach extends \textsc{FLowHigh} ~\cite{flowhigh}, a conditional flow matching (\textsc{cfm}) model, in two ways: (i) we apply the model, originally proposed for audio speech super-resolution, to the music domain; (ii) we extend the model with classifier-free guidance and audio feature controls to have fine-grained control over the restoration output.

\textsc{FLowHigh} offers efficient audio bandwidth extension with a key advantage of single-step inference, leading to comparatively faster inference than diffusion models.
\textsc{FLowHigh} utilizes a transformer-based vector field estimator (35.4M parameters: 2 layers, 16-head self-attention, 1024 embedding dim, 4096 dimension for feed-forward networks). 
Given a mel-spectrogram of the narrow band (i.e. degraded) signal, the model predicts a full band mel-spectrogram from which the audio waveform is reconstructed using a pre-trained and frozen \textsc{BigVGAN} vocoder~\cite{lee2022bigvgan}. 
Following the original \textsc{FLowHigh} method, and in order to minimise artefacts, a post-processing step copies the input signal up to a cutoff frequency into the output. As a result, only the frequencies that were not already present in the input are used from  the model prediction. In this work, we consider both \textsc{cfm}strategies proposed by \textsc{FLowHigh}: (i) \textit{Adaptive} CFM, which uses a linear interpolation path from $\mathbf{x}_0$ to $\mathbf{x}_1$, coupled with an adaptive noise schedule; (ii) \textit{Mixed} CFM, which employs a similar path from $\mathbf{x}_0$ to $\mathbf{x}_1$ for the lower frequencies, and another path from Gaussian noise to $\mathbf{x}_1$ for the higher frequencies. We refer the readers to Section III of the original \textsc{FLowHigh} paper and code\footnote{\url{https://github.com/jjunak-yun/FLowHigh\_code}} for more details.

\subsection{Controllability}
We modify the baseline \textsc{FlowHigh} model by allowing fine-grained controllable audio restoration based on audio features, such as spectral centroid and rolloff~\cite{mcfee2015librosa}. We achieve this by leveraging \textsc{cfg-zero$\star$} \cite{cfg_zero}, an improved guidance strategy for flow matching on top of the usual Classifier-Free Guidance (\textsc{cfg}), based on the following equation: $\hat{\mathbf{v}}(\mathbf{x}_t, t, \mathbf{c}) = (1 - w) \cdot s \cdot \mathbf{v}_{\theta}(\mathbf{x}_t, t, \emptyset) + w \cdot \mathbf{v}_{\theta}(\mathbf{x_t}, t, \mathbf{c})$. An adaptive scale, $s$, is introduced to dynamically adjust the magnitude of the unconditional velocity field, ensuring that it aligns more accurately with its conditional counterpart. The optimal scale $s^\star$ can be obtained by the following closed-form solution, where $s^\star$ is the projection of the conditional velocity onto the unconditional one:

\begin{equation} s^\star = \frac{\langle \mathbf{v}_{\theta}(\mathbf{x}_t, t, \mathbf{c}), \mathbf{v}_{\theta}(\mathbf{x}_t, t, \emptyset) \rangle}{|\mathbf{v}_{\theta}(\mathbf{x}_t, t, \emptyset)|^2_{2}} \label{eq:s_star} \end{equation}

In our implementation, the conditioning vector $\mathbf{c} \in \mathbb{R}^{m \times F}$ consists of $m$ distinct audio features extracted over $F$ frames. These features (e.g. spectral rolloff) provide fine-grained control over the restoration process, which enables our approach to generate audio that adheres closely to the desired acoustic characteristics.

\subsection{Audio Feature Control: Dynamic Spectral Contour}
\label{sec:dsc}

\begin{figure}[t]
    \centering
    \includegraphics[width=\linewidth,trim={0.2cm 0.1cm 1.5cm 0.5cm},clip]{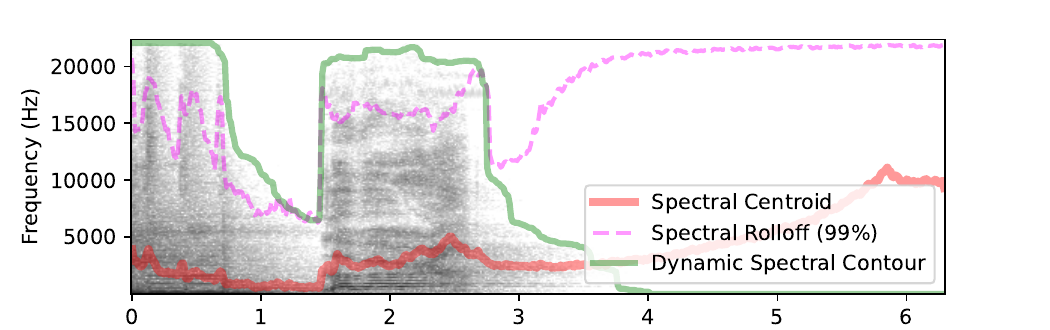}
    \caption{Comparison of bandwidth control features. Our proposed Dynamic Spectral Contour accurately captures bandwidth where spectral centroid and rolloff fail (e.g., in the low-energy region from 4s onwards).
    }
    \label{fig:features}
\end{figure}

Empirical observation revealed that standard signal features such as the spectral centroid and rolloff can take large values in silent regions, which we hypothesise may be problematic to provide an intuitive control signal - see Figure \ref{fig:features}.
We introduce a novel time-dependent feature which we call  Dynamic Spectral Contour (\textsc{dsc}). Intuitively, this feature can be thought of as the highest meaningfully active frequency bin in the spectrum, so that it is a proxy representation of the bandwidth. 
Conditioning the model on the \textsc{dsc} allows the user to define the frequency up to which they want the model to restore content, therefore allowing minute adjustments specific to the needs of each use case.

The computation of the \textsc{dsc} is a multi-step process designed to robustly track the spectral edge of the signal over time. First, we compute the short-time Fourier transform (\textsc{stft}) (n\_fft=2048, hop length=512) of the input audio to obtain its magnitude spectrogram. A binary mask is then created by thresholding the spectrogram magnitudes against a value $q$. To smooth this mask along the frequency axis, we apply a 1D Gaussian filter with a standard deviation of $\sigma_f$. For each time frame, we then find the first frequency bin where the value of this smoothed mask drops below a threshold $\gamma$. This frequency, which represents the spectral edge for that frame, forms the initial contour. Finally, to ensure temporal smoothness, a median filter with a window size of $m_f$ is applied to this sequence of frequency values. For our experiments, we use the following hyperparameters: $q=10^{-1.6}$, $\sigma_f=9$, $\gamma=0.07$, and $m_f=9$, which we empirically found to create a good `spectral topline'.


\section{Experiments and Results}
\label{sec:results}



\subsection{Dataset and preprocessing}
\label{sec:data}

Our dataset consists of 8,503 tracks (425h) from a commercial music catalog, resampled to 44.1 kHz and segmented into 1.5s clips, with an 8:1:1 train-validation-test split.

We train for bandwidth extension using an on-the-fly data augmentation scheme where each clean sample is paired with a degraded version. The degradation is performed by applying a low-pass filter randomly selected from four families: Finite Impulse Response (\textsc{fir}), Biquad, Chebyshev Type I, and an ideal brick-wall filter. To ensure robustness, filter parameters such as order, ripple, and a cutoff frequency (sampled from 3-18 \unit{\kilo\hertz} with 1\unit{\kilo\hertz} intervals) are also randomized, creating a diverse set of over 400 unique filters for training.


\subsection{Evaluation}
\label{sec:evaluation}
We use audio metrics \textsc{fad}\textsubscript{CLAP}, log-spectral distance (\textsc{lsd}) \cite{rabiner1993fundamentals} and Log Kurtosis Ratio Perceptual Indicator (\textsc{lkr-pi}) to evaluate our method.
The (\textsc{fad}) is a reference-free metric that evaluates the quality of generated audio by measuring the statistical distance between distributions of real and synthetic samples in a deep embedding space~\cite{roblek2019fr}. We use audio embeddings from \textsc{laion-clap}\cite{laionclap2023} as it takes in 48kHz audio, which preserves the full bandwidth for evaluation.  We compute the \textsc{fad} scores using an open-source toolkit\footnote{\url{github.com/gudgud96/frechet-audio-distance}}.
The (\textsc{lsd}) is a widely-used objective metric for quantifying the spectral dissimilarity between a reference signal and an estimated signal. It is computed as the root mean square error (RMSE) between their log-magnitude spectra, providing a perceptually relevant measure of distortion, expressed in decibels (dB).
The \textsc{lkr-pi}~\cite{torcoli2019improved} metric quantifies perceived musical artifacts arising from spectral holes by measuring the change in spectral kurtosis after processing. 
A score close to zero suggests that the processed signal has successfully preserved the harmonic and tonal structure of the original.

To measure control adherence, we compute the absolute log-distance between the input control signal $c$, and the control signal extracted from the restored audio $\hat{c}$ with $|\log(\hat{c}) - \log(c)|$. Using the log-scale preserves the perceptual relationship between frequencies, and provides a relative error between the base and the reconstructed signal.
We report the median distance, as it offers a more robust measure of central tendency by being less sensitive to outliers.

\begin{table}[t]
 	\centering
 	\scriptsize 
 	\setlength{\tabcolsep}{4pt}
    \resizebox{\columnwidth}{!}{
    \begin{tabular}{lcc|cc|cc}
 	~ & ~ & ~ & \multicolumn{2}{c|}{$\mathbf{f_c = 4 \text{ \unit{\kilo\hertz}}}$} & \multicolumn{2}{c}{$\mathbf{f_c = 8 \text{ \unit{\kilo\hertz}}}$} \\
 	\textbf{Method} & ~ & steps & $\text{FAD}_{\text{CLAP}}$ $\downarrow$ & LSD $\downarrow$ & $\text{FAD}_{\text{CLAP}}$ $\downarrow$ & LSD $\downarrow$ \\
 	\midrule
 	\textsc{1d-diff}\cite{vrdmg} & \textsc{dc+rg} & $35$ & 0.23 & 2.25\stddev{0.64} & 0.07 & 1.64\stddev{0.54} \\
 	\textsc{1d-diff}\cite{vrdmg} & \textsc{pigdm}  & $35$ & 0.25 & 2.31\stddev{0.57} & 0.12 & 1.78\stddev{0.47} \\
 	\textsc{cqt-diff}\cite{cqtdiff} & \textsc{dc+rg}  & $35$ & 0.49 & 3.52\stddev{0.99} & 0.18 & 2.21\stddev{0.71} \\
 

    \rowcolor{gray!20} \textsc{FlowHigh} & \textsc{adaptive} & 1 & 0.15 & 1.58\stddev{1.01} & 0.10 & 1.50\stddev{1.10} \\
         
 

    \rowcolor{gray!20} \textsc{FlowHigh} & \textsc{mixed} & 1 & 0.18 & 1.55\stddev{0.99} & 0.12 & 1.42\stddev{1.02} \\


    \end{tabular}
    }
 	\caption{Results for bandwidth extension without using control signals. Overall top performing model in gray.
    }
 	\label{tab:results_sota}
\end{table}

\begin{figure*}
    \centering
    \includegraphics[width=\linewidth,trim={1cm 1.2cm 1cm 1cm},clip]{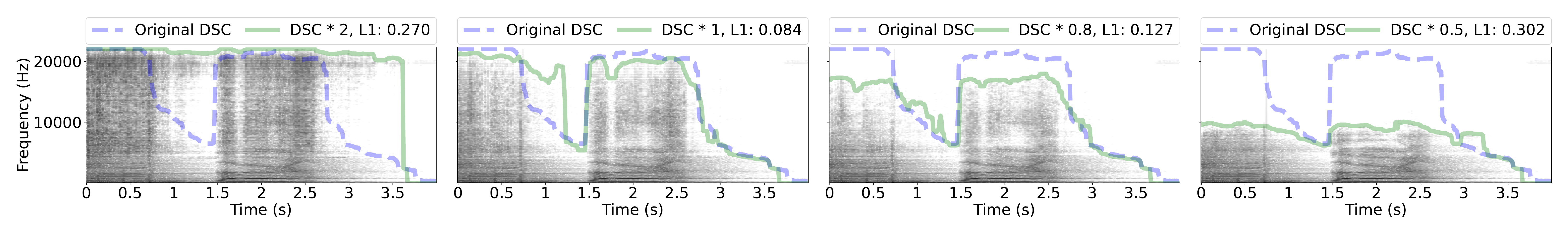}
    \caption{
    Audio restoration effect of manipulating \textsc{dsc} control signals and a 4kHz degraded input.
    In purple, the \textsc{dsc} calculated from the original clean file. 
    In green, \textsc{dsc} calculated from the restored audio using the manipulated \textsc{dsc} as control signal.
    }
    \label{fig:placeholder}
\end{figure*}

\subsection{Experiment 1: Full-band Restoration}
\label{sec:exp1_uncontrolled}

To evaluate the performance of our proposed model, we conducted comparative experiments against two diffusion-based baseline models: \textsc{cqt-diff} and \textsc{1d-diff}. \textsc{cqt-diff} is a generative model that operates in the Constant Q Transform (\textsc{cqt}) domain. It employs a U-Net architecture trained only on clean audio data, enabling it to leverage the rich time-frequency representation of the \textsc{cqt} for audio generation and restoration. In contrast, the second baseline, \textsc{1d-diff}, is a diffusion model that directly processes raw audio waveforms. It utilizes a 1D U-Net architecture and offers a direct, end-to-end approach that foregoes spectral transformations.

\subsubsection{Results}
\label{sec:exp1_results}

Table \ref{tab:results_sota} presents the evaluation of our proposed  model against the \textsc{cqt-diff} and \textsc{1d-diff} baselines for the task of bandwidth extension. The results indicate that, overall, our approach outperforms both across all tested conditions. In the more challenging 4 \unit{\kilo\hertz} cutoff scenario, the Mixed \textsc{cfm}variant secures the lowest Log-Spectral Distance (\textsc{lsd}=1.55) and the Adaptive \textsc{cfm}has the lowest Fréchet Audio Distance (\textsc{fad}=0.15). Both results are substantially better than the best-performing baseline, \textsc{1d-diff} (\textsc{fad}=0.23, \textsc{lsd}=2.25). 
In the 8 kHz cutoff condition, \textsc{FLowHigh} maintains a strong performance. Specifically, both of our flow-based methods attain a lower \textsc{lsd} (1.50 \& 1.42) compared to the best \textsc{1d-diff} baseline (1.64). This demonstrates that \textsc{FLowHigh} is not only more effective but also more robust for the task of audio bandwidth extension.

\subsection{Experiment 2: Controlled Restoration}
\label{sec:exp2_controlled}

In our second experiment, we investigate the model's ability to restore audio that adheres to a given fine-grained acoustic feature. We train separate instances of the model using three types of control signals: our proposed \textsc{dsc}, spectral centroid and rolloff. 

\subsubsection{Results}
\label{sec:exp2_results}

Table \ref{tab:metrics_dsc} shows the results for single-step bandwidth extension on a 4kHz degraded audio. The control signal used is extracted from the ground truth clean audio. We observe that when using the spectral centroid as the control signal, the reconstruction metrics and control adherence error is the worst. We interpret this as a consequence of the numerical values of spectral centroid not aligning well with the notion of bandwidth in quiet and silent parts, as illustrated in Figure \ref{fig:features}. We find our proposed \textsc{dsc} to be the best performing feature, both in terms of reconstruction and control adherence. Compared to Table \ref{tab:results_sota}, we notice comparable results in \textsc{fad} and improved results for \textsc{lsd} when control signals are introduced.

\begin{table}[t]
\centering
\resizebox{\columnwidth}{!}{%
\begin{tabular}{l cccc c}
~ & \multicolumn{4}{c}{~} & \multicolumn{1}{c}{Control} \\
~ & \multicolumn{4}{c}{Reconstruction Metrics} & \multicolumn{1}{c}{Adherence} \\
\cmidrule(lr){2-5} \cmidrule(lr){6-6}
\multicolumn{1}{c}{Control signal} & \textbf{FAD\textsubscript{CLAP}} $\downarrow$ & \textbf{LSD} $\downarrow$ & \textbf{LKR-PI} $\downarrow$ & \textbf{MSE\textsubscript{MFCC}} $\downarrow$ & \textbf{Abs. Log Distance} $\downarrow$ \\

\midrule


\rowcolor{gray!0} \multicolumn{6}{l}{\textbf{Only conditioning ($w=1$)}} \\
\quad Centroid & 0.41 & 4.04\stddev{0.98} & -0.70\stddev{0.36} & 22.91\stddev{20.86} & 1.41 \\
\quad Roll-off & 0.19 & 1.69\stddev{0.97} & 0.00\stddev{0.15} & 5.60\stddev{6.30} & 0.30 \\
\rowcolor{gray!10} \quad \textsc{dsc} & 0.12 & 0.99\stddev{0.29} & -0.06\stddev{0.22} & 4.83\stddev{8.27} & 0.18 \\

\rowcolor{gray!0} \multicolumn{6}{l}{\textbf{Guidance ($w=3$)}} \\
\quad Centroid & 0.40 & 3.31\stddev{0.85} & -0.36\stddev{0.28} & 43.56\stddev{32.30} & 0.93\\
\quad Roll-off & 0.21 & 1.76\stddev{0.92} & -0.09\stddev{0.20} & 9.03\stddev{8.81} & 0.38 \\
\rowcolor{gray!10} \quad \textsc{dsc} & 0.14 & 1.05\stddev{0.29} & -0.06\stddev{0.20} & 6.07\stddev{8.38} & 0.24 \\
\end{tabular}
}
\caption{The \textsc{dsc} feature overall produces the best results.}
\label{tab:metrics_dsc}
\end{table}

\vspace{-0.2cm}
\subsubsection{Manipulating Dynamic Spectral Contour}
\label{sec:exp2_results}
We further evaluated control adherence using the \textsc{dsc}. By scaling the \textsc{dsc} by factors of 0.5 and 2.0 to guide restoration, we simulate a user controlling the desired bandwidth of the restored signal. A factor of 1 meaning full original bandwidth, 0.5 half bandwidth etc. We also include a test condition with a factor of 2 to test the behaviour of the model when pushed to extend the bandwidth beyond natural bounds. 
Figure \ref{fig:placeholder} shows an illustration of the effect of manipulating the \textsc{dsc} control signal for various values of the multiplicative factor. 

The results in Table \ref{tab:results_4k_revised} for a factor 0.5 show that the adherence to the control signal remains reasonable while the \textsc{fad} does not degrade significantly as compared with values in Table \ref{tab:metrics_dsc}. This indicates respectively that: (1)  following a scaled control signal is slightly more difficult than reconstructing to the ground truth bandwidth, and (2) the reconstructed audio distribution remains within comparable distance to the ground truth distribution, which suggests that at most a minor degradation of audio quality may be introduced by scaling the control signal. 
In the case of factor 2, the adherence to the control signal degrades significantly, and the \textsc{fad} also tends to be affected. Our interpretation is that this condition pushes the model to generate a signal that is overall close to the Nyquist frequency (cf. Figure \ref{fig:placeholder}, far left), which is an unnatural audio profile and therefore leads to the introduction of significant artifacts. 
We conclude from these results that the model appears to be able to produce meaningful bandwidth reconstructions so long as the \textsc{dsc} control remains within a range that reflects a plausible audio signal.

\begin{table}[t]
    \centering
    \scriptsize
    \setlength{\tabcolsep}{4pt}
    \resizebox{0.95\columnwidth}{!}{%
        \begin{tabular}{llcc}
            \textbf{Control signal} & \textbf{FAD\textsubscript{CLAP}} $\downarrow$ & \textbf{Multiplication factor} & \textbf{Abs. Log Distance} $\downarrow$ \\
            \midrule
            \rowcolor{gray!0} \multicolumn{3}{l}{\textbf{Only conditioning ($w=1$)}} \\
            \quad \textsc{dsc} & 0.12 & 0.5 & 0.46 \\
            \quad \textsc{dsc} & 0.13 & 2 & 1.11 \\
            \rowcolor{gray!0} \multicolumn{3}{l}{\textbf{Guidance ($w=3$)}} \\
            \quad \textsc{dsc} & 0.16 & 0.5 & 0.35 \\
            \quad \textsc{dsc} & 0.23 & 2 & 9.21 \\
        \end{tabular}
    }
\caption{Bandwidth extension results for different \textsc{cfg} weightings and factors. All models trained with Mixed CFM. 
}
\label{tab:results_4k_revised}
\end{table}


\vspace{-0.2cm}
\section{Conclusion}
\label{sec:conclusions}

This paper presented a novel and efficient approach to music restoration that integrates flow matching with fine-grained controllability. By leveraging the strengths of conditional flow matching and a transformer-based vector field estimator, our approach reconstructs full band audio from degraded narrow band inputs with a single-step sampling process. Furthermore, we introduced the Dynamic Spectral Contour (\textsc{dsc}) as a novel audio feature to guide the restoration process, specifically enabling to control the extent to which the bandwidth is extended.
It is our hope that the type of fine-grained controllability offered by features like \textsc{dsc} opens avenues for the emergence of tools offering a meaningful interactivity for creatives. Opportunities for future work include further improving audio quality, and exploring the broader application of fine-grained controllable flow matching in various audio manipulation tasks.

\bibliographystyle{IEEEbib}
\bibliography{refs}

\end{document}